\documentstyle[preprint,aps,psfig,amssymb]{revtex}
\begin{document}
\title{Internal energy of high density Hydrogen: Analytical
approximations compared with path integral Monte Carlo calculations}
\author{S.A.~Trigger$^1$, W.~Ebeling$^1$, V.S.~Filinov$^2$, V.E.~Fortov$^2$, and 
M.~Bonitz$^3$}

\address{$^1$Institut f\"ur Physik, Humboldt-Universit{\"a}t Berlin\\
Invalidenstrasse 110 D-10115 Berlin}
\address{$^2$Russian Academy of Sciences,
Institute for High Energy Density,
Izhorskaya street 13-19, Moscow 127412, Russia}
\address{$^3$Fachbereich Physik, Universit{\"a}t Rostock\\
Universit{\"a}tsplatz 3, D-18051 Rostock, Germany}
\date{\today}
\maketitle
%----------------------------------------------------------------------------------------------------
\begin{abstract}
The internal energy of high-density hydrogen plasmas in the temperature range
$T = 10,000 \dots 50,000 K$ is calculated 
by two different analytical approximation schemes
(method of effective ion-ion interaction potential - EIIP and Pad\'e approach 
within the chemical picture - PACH) 
and compared with path integral Monte Carlo results. 
Reasonable agreement between the results obtained from the three 
independent calculations is found, the reasons for still existing
differences is investigated. Interesting
high density phenomena such as the  formation of clusters and the onset of
crystallization are discussed.
 
\end{abstract}
%------------------------------------------------------------------
\pacs{  }
%-------------------------------------------------------------------
%-------------------------------------------------------------------
%\newpage
\section{Introduction}\label{intro}

The thermodynamics of strongly correlated Fermi systems at high pressure
are of growing importance in many fields, including shock and laser plasmas, 
astrophysics, solids and nuclear matter, see
Refs.~\cite{boston97,binz96,green-book,bonitz-book,Haberland} for an overview.
In particular, the thermodynamic properties of hot dense plasmas 
are essential for the description 
of plasmas generated by strong lasers \cite{Haberland}.
Further, among the phenomena of current interest are the high-pressure 
compressibility of deuterium \cite{dasilva-etal.97}, metallization
of hydrogen \cite{weir-etal.96}, plasma phase transition etc., 
which occur in situations where both
{\em interaction and quantum effects} are relevant. 
Among the early theoretical papers on dense hydrogen we refer to
Wigner/Huntington \cite{Wigner}, Abrikosov \cite{Abrikossov}, 
Ashcroft \cite{Ashcroft} and Brovman et al. \cite{Brovman} and, concerning 
the plasma phase transition, see Norman and Starostin \cite{NormanStarostin},
Kremp et al. \cite{kremp}, 
Saumon and Chabrier \cite{saumon} and Schlanges et al. \cite{schlanges},
as well as to some earlier
investigations of one of us \cite{red-book,Eb85,Pade85,BeEb99}. Among the early simulation 
approaches we refer to several Monte Carlo (MC) calculations, e.g. 
\cite{Zamalin,Hansen,DeWitt}.
%\\

There has been significant
progress in recent years in studying these systems analytically and numerically, see e.g.
\cite{boston97,binz96,bonitz-book,KKT90,bonitz-etal.96jpcm,vova01,EbSt99,KTR94} 
for an overview. However, there remains an urgent need
to test analytical models by an independent numerical approach.
Besides the molecular dynamics approach, e.g. \cite{KKT90,vova01}, the
path integral Monte Carlo (PIMC) method 
is particularly well suited to describe thermodynamic
properties in the region of high density. This is because it starts from 
the fundamental plasma particles - electrons and ions,
(physical picture) and treats all interactions, including bound 
state formation, rigorously and selfconsistently.
We notice remarkable recent progress in applying these techniques to  Fermi systems,
for an overview see e.g. Refs.~\cite{boston97,binz96,binder96,berne98}.

Several methods have been developed to perform quantum MC. First we mention
the restricted PIMC method 
\cite{ceperley95,ceperley95rmp,mil-pol,militzer-etal.00};
here special assumptions on the
density operator ${\hat \rho}$ are introduced in order to reduce the sum over
permutations to even (positive) contributions only. It can be shown, however,
that this method does not reproduce the correct ideal Fermi gas limit \cite{Filinov01}.
An alternative are direct fermionic PIMC simulations which have occasionally
been attempted by various groups.
Recently, three of us have proposed a new path integral representation
for the N-particle density operator
\cite{filinov-etal.KBT,filinov-etal.99xxx1,filinov-etal.00jetpl,filinov-etal.00pla}
which  allows for {\em direct fermionic path integral Monte Carlo} simulations
of dense plasmas in a wide range of densities and temperatures. Using this
concept, the pressure and energy of a degenerate strongly coupled
hydrogen plasma have been computed
\cite{filinov-etal.99xxx1,filinov-etal.00jetpl,filinov-etal.00pla,FiBoEbFo01} 
as well as the pair distribution functions in the region of partial 
ionization and dissociation \cite{filinov-etal.00jetpl,filinov-etal.00pla}. 
This scheme is
rather efficient when the number of time slices (beads) in the path integral
is less or equal 50 and was found to
work well for temperatures $k_BT > 0.1Ry$. 

One difficulty of PIMC simulations is that reliable error estimates are
often not available, in particular for strongly coupled degenerate systems.
Here, we will make a comparison with two independent analytical methods.
The first is the method of an effective ion-ion interaction potential (EIIP) 
which has previously been developed for application to simple solid and 
liquid metals \cite{Brovman,KKT90} and which is here, for the first time 
adopted to dense hydrogen. The second is the
method of Pad\'e approximations in combination with Saha equations, i.e. the 
chemical picture (PACH)
\cite{green-book}. The Pad\'e formulas are constructed on the basis of the
known analytical limits of  low density \cite{green-book,ebeling} and
high density \cite{green-book}, and they are exact up to
quadratic terms in the density, interpolating between
the virial expansions and the high-density asymptotics
\cite{Pade85,Pade90,EbFo91}. 

We will show here that both methods, EIIP and PACH provide results for the internal energy
which agree well with each other at high densities where the electrons are 
strongly degenerate and no bound states exist, 
approximately for $n>10^{24}$cm$^{-3}$. In this region, 
there is also good agreement with recent density functional results 
\cite{Xu98}.
The agreement with the PIMC results is very good below $10^{22}$cm$^{-3}$. 
For intermediate densities, where the degree of ionization changes strongly,
we observe deviations. Also, at high densities, the PIMC results,
tend to lower energies than the analytical approaches. Finally, they
reveal several interesting effects, such 
as the formation of clusters and the onset of ion crystallization.

\section{Physical parameters and basic effects}\label{par}

Let us study a hydrogen plasma consisting of $N_e$ electrons and $N_p$
protons ($N_e = N_p$). The total proton (atom) density is 
$n = N_p/V$. 
The average distance between the electrons is the Wigner-Seitz radius 
$d = [3/4\pi n]^{1/3}$, and other characteristic lengths are 
the Bohr radius $a_{\rm B} = \hbar /me^2$,
the Landau length $l = e^2 / kT$ and the 
De Broglie wave length $\Lambda_{\rm e} = h/[2\pi m_{\rm e} kT]^{1/2}$ 
of the electrons. The degeneracy parameter is $n \Lambda_{\rm e} ^3$.
We define the dimensionless temperature $\tau = kT/Ry$ 
which, in the considered below temperature interval, varies between $0.06< \tau <0.4$. 
Furthermore, we introduce the Wigner-Seitz parameter
$r_s = d/a_B$ and the dimensionless classical coupling strength
$\Gamma = e^2/(kT d)$.

Hydrogen is anti-symmetric with respect to the charges ($e_-=-e_+$) and 
symmetric with respect to the densities ($n_+=n_-=n$) and, due to the big 
mass difference,  $m_p = 1836~m_{\rm e}$,
ions and electrons behave quite differently. At the considered temperatures,
 the ions may be treated classically as long as $n \lesssim 10^{27}$ cm$^{-3}$.
Further, for these temperatures and densities, the proton coupling parameter is 
in the range $0< \Gamma < 150$, i.e. we expect strong 
coupling effects.
We study in this work the internal energies of the fluid hydrogen system and 
start with providing some simple estimates for guidance.
In the following we will give all energies in Rydberg units. 

First, at very low densities the electrons and the protons behave like
an ideal Boltzmann gas. Therefore, the energy per proton 
(of free electrons and protons) is given by
 (in Rydberg units) 
\begin{equation}
\epsilon = E / N = 3 \tau.
\end{equation}
In other words the low-density limit is, in our temperature interval, 
a positive 
number in the region $\epsilon \simeq 0.2 - 1.2$. With increasing density we
expect a region where atoms and, possibly also a few molecules, are formed
\cite{red-book,filinov-etal.00pla}. In the region of atoms a lower bound for 
the energy per proton is
\begin{equation}
\epsilon = \frac{3}{2} \tau - 1,
\end{equation}
where the last term represents the binding energy $1 Ry$ of H-atoms. 
If molecules are formed, 
the corresponding estimate per proton is still lower
\begin{equation}
\epsilon = \frac{3}{4} \tau - 1.17.
\end{equation}

Generally, the existence of a lower bound for the energy per proton 
was proven by Dyson and Lenard \cite{dyson-etal.67} and Lieb and Thirring \cite{lieb-etal.75}
\begin{equation}
E/N > - C,
\end{equation}
where the best estimate known to us (which certainly is much too large), is $C \simeq 23$ 
\cite{lieb-etal.75}.
We see that, with increasing density,
the energy per proton tends to negative values and may reach a finite
minimum. Further density increase will cause the energy to increase again 
as a result of quantum degeneracy effcts. 

In order to understand this increase let us look first at the limit of very
high density (still in the region where the protons are classical). Then the
first estimate of the energy is
\begin{equation}
\epsilon = \frac{3}{2} \tau + \frac{2.21}{r_s^2},
\label{fermi}
\end{equation}
which is positive.
The last term, representing the Fermi energy of the electrons, is strongly 
increasing with density (with power $n^{2/3}$). 
In the next approximation  according to 
Wigner's estimate we have to take into account the Hartree 
contribution to the electron energy and a corresponding estimate for the proton
energy. The proton energy is estimated under the assumption that protons 
form a lattice. This way we find the estimate 
\begin{equation}
\epsilon = \left(\frac{3}{2} \tau - \frac{1.793}{r_s }\right) + 
\left(\frac{2.21}{r_s^2} - \frac{0.916}{r_s} \right).  
\label{estimate}
\end{equation}
The two corrections that were added to Eq. (\ref{fermi}) are both negative and 
scale like $n^{1/3}$. In other words, these interaction terms might play a major
role with decreasing density.
At a critical density the energy per proton may become negative. This densitiy 
can be estimated from Eq.~(\ref{estimate}) by
solving the quadratic equation 
\begin{equation}
0 = \frac{3}{2} \tau \,r_s^2 - 2.709 \,r_s  + 2.21,
\end{equation}
perturbatively, starting with the zero temperature limit, and adding the 
first (linear in $\tau$) correction,
\begin{equation}
r_s^0 \simeq 0.816 + 0.37 \tau  + ... .
\label{rs0}
\end{equation}
This result coincides, for $\tau \rightarrow 0$, with Wigner's criterion for 
the existence of molecules: for $d < a_B$, molecules cannot exist
since there is no room for forming bound state wave functions. According to 
Eq.~(\ref{rs0}), for finite 
temperature, molecules exist only for still larger $d$ as thermal 
fluctuations increase the wave function overlap. More generally, with 
increasing temperature, the energy becomes positive at lower 
density compared to the case $T=0$.

Summarizing the qualitative results obtained in this section we may state
that we expect, in the given temperature range,
the following general behavior of the internal energy per proton:
at zero density the energy starts with the ideal gas expression which depends only
on the temperature. With increasing density the energy per proton becomes negative
due to correlation effects (bound states, electron correlations, proton correlations).
A minimum is formed and at a density where the proton density is close to 
the inverse Bohr radius cubed the energy per proton turns to positive values and is more and
more determined by the ideal electron energy increasing with $n^{2/3}$, corrected by 
correlation contributions of order $n^{1/3}$ which are determined by the 
Hartree term and by proton-proton coupling effects.
In the following we will show that this qualitative picture is 
supported by the results of our calculations.

\section{Method of an effective ion-ion interaction potential}\label{eiip}
It is well known that in plasmas and plasma-like systems, in a broad parameter range,  
the interaction between the electron and ion subsystems is  
weak, whereas the interacrtions within the electron and ion subsystems can be strong. 
The corresponding small parameter is the ratio $u_{ei}/{E_F}$ of the characteristic 
value of electron--ion interaction $u_{ei}$ to the electron Fermi energy $E_F$.
Therefore, the mentioned approximation is valid for systems with 
degenerate electrons, if $E_F \gg T_e \ge T_i$, where $T_e$ and $T_i$ 
are the electron and ion temperatures respectively (below we will consider 
the case $T_e = T_i$). 
Typical systems for which this approximation is fulfilled are simple solid and 
liquid metals and non-transitional metals in general, and this approximation 
serves as the basis for the computation of thermodynamic and electron kinetic 
properties, e.g. \cite{KKT90,Harrison}. 
 
For simple metals the Fermi energy is not very large compared to  
the characteristic electron--ion Coulomb interaction, taken at the average  
interparticle distance. However, due to the orthogonality of the wave  
functions for the conduction electrons and electrons bound in the ion shells, 
there is a partial compensation of the electron--ion Coulomb attraction at
small distances which effectively weakens the electron--ion interaction. 
This fact is described in the theory of simple metals in the framework of  
the so called pseudopotential  theory. The calculation of the  
pseudopotential is, in general, a complicated problem in particular due to
its   non-local structure \cite{Harrison,BoTr}.  For practical applications
it can be  represented approximately as a local interaction with one or two 
fitting parameters for each metal. On basis of the pseudopotential 
theory all thermodynamic properties and electronic kinetic coefficients 
can be calculated with sufficiently high accuracy for a wide range of  
temperatures and pressures. Naturally, these calculations require
reliable knowledge of the properties of the
two quasi--independent subsystems: the degenerate electron liquid on 
the positive charge background and the classical ion subsystem with some 
effective strong inter-ion interaction. 
  
It is apparent that there is also a wide range of parameters for 
highly ionized strongly compressed hydrogen plasmas, where the electron--ion
interaction  is weak. For these parameters  
the complicated problem of calculating the properties of
a strongly coupled quantum electron--proton system can be  
essentially simplified. 
In so doing, the results obtained for high compression (when no
bound electron states -- hydrogen atoms and molecules -- 
are existing), do not require any fitting, 
in contrast to the case of simple metals, because the inter--ion 
potential for hydrogen is pure Coulomb.  
Therefore, the data obtained with this analytical approximation,  
can be considered as an reliable basis for comparison with the results of
alternative approaches, including analytical and simulation methods for 
degenerate quantum systems of Fermi particles.
The results of this pseudopotential approach are especially important for 
conditions of extreme compression where the plasma is characterized by 
strong interaction within the electron and, especially, the ion subsystem.
For these difficult situations experimental data are still missing whereas  
new acurate numerical methods for Fermi system are only emerging. 
 
Let us consider the Hamiltonian of an electron--proton plasma, for which 
the terms with infinite zero-components of the potentials are canceled, 
due to quasineutrality (for generality we retain the charge number $Z$ of the ions): 
\begin{eqnarray}\label{eq3.1} 
H&=& \sum_k\epsilon_ka^\dagger_{k} a_k+\frac{1}{2V} \sum_{k,k',q \neq 0}  
\frac{4\pi e^2}{q^2} a^\dagger_{k-q} a^\dagger_{k'+q} a_{k'} a_k  
+ \frac{1}{V} \sum_{k,q'\neq 0}u_{ei}(q)a^\dagger_k a_{k+q} 
 \sum_{j=1}^{N_i} e^{i\vec{q}\vec{R}_j} 
\nonumber\\ 
&+& \frac{1}{2V} \sum_{i\neq j} \sum_{q\neq 0} \frac{4\pi Z^2e^2}{q^2}  
e^{i\vec{q} \left(\vec{R}_i-\vec{R}_j\right)}+K_i. 
\end{eqnarray}  
Here $\epsilon_k$ is the energy of the electron with momentum $\hbar k$ and  
$u_{ei}(q)=-\frac{4\pi Ze^2}{q^2}$  
is the Fourier-component of the electron--proton interaction potential. 
For the electron degrees of freedom in the Hamiltonian $H$ the representation 
of second quantization is used where $a^\dagger_p$ and $a_p$ are, respectively, 
the operators 
of creation and annihilation of an electron with momentum $p$ .  
For the classical ions the coordinate representation is more convenient, thus 
in Eq.~(\ref{eq3.1}) $R_i$ denotes the coordinate of the i-th ion. 
To calculate the plasma energy, as in the theory of simple metals 
\cite{Brovman,KKT90}, 
two  main approximations have to be used. The first is the adiabatic 
approximation for the ion motion, which is slow compared to the electron one. 
The second is the smallness of the ratio of the characteristic electron--proton
Coulomb interaction to the Fermi energy $E_F$. The respective parameter is 
 $\displaystyle{\Gamma_{ei}=\frac{Z e^2}{dE_F}=Z\Gamma \frac{kT}{E_F} \sim
n^{-1/3}}$.
Calculation of the electron energy in the external field of the immobile 
ions (protons) leads to the energy of the plasma given as function of the ion 
coordinates $R_j$. In general, the perturbation theory in terms of the 
parameter $\Gamma_{ei}$ 
gives rise not only to pair but, naturally, also to higher order ion--ion 
interactions, which are rather complicated.  To second order of perturbation 
theory in the parameter $\Gamma_{ei}$ the energy per one electron of a plasma with a 
fixed proton configuration $\left\{R_j\right\}$ is easily written,    

\begin{eqnarray}\label{eq3.2} 
\frac{E\left(\left\{R_j\right\}\right)}{N_i}&=& \frac{\langle H\rangle_e}{N_i} = 
\epsilon_e + \frac{3}{2} kT - \frac{1}{2} \int  
\frac{d^3q}{(2\pi)^3} \frac{u_{ei}^2(q)\Pi_e(q)}{\varepsilon_e(q)} 
\nonumber\\ 
&-& \frac{Z^2 n_i}{2\Pi_e(q=0)} + \frac{1}{2VN_i} \sum_q \sum_{i\neq j} {\cal{V}}^{\rm{eff}}(q) 
e^{i\vec{q}(\vec{R}_i-\vec{R}_j)}. 
\end{eqnarray} 
Here $\epsilon_e$ is the energy (per ion) of the {\em correlated electron liquid} on the  
homogeneous positive charge background. The functions $\Pi_e(q)$ and
$\varepsilon_e(q)$  are,  respectively, the static polarization function and
the static dielectric function   of the correlated electron liquid. These
functions are related to one another by the  usual equality:  
                                                               
\begin{equation}\label{eq3.3} 
\varepsilon_e(q)=1+ \frac{4\pi e^2}{q^2} \Pi_e(q). 
\end{equation}                                                                
The Fourier-component of the effective pair interaction potential 
between the ions, ${\cal{V}}_{ii}^{\rm {eff}}$, which appears in (\ref{eq3.2}) 
has the form:  
                                                                        
\begin{equation}\label{eq3.4} 
{\cal{V}}_{ii}^{\rm{eff}}(q)= \frac{4\pi Z^2 e^2}{q^2} - u_{ei}^2(q) 
\frac{\Pi_e(q)}{\varepsilon_e(q)} =  
\frac{4\pi Z^2 e^2}{q^2\varepsilon_e(q)}. 
\end{equation} 
In the following, we will concentrate on hydrogen and set $Z=1$ leading to 
the effective proton-proton interaction
\begin{equation}\label{eq3.4h} 
{\cal{V}}_{pp}^{\rm{eff}}(q)= \frac{4\pi e^2}{q^2\varepsilon_e(q)}. 
\end{equation} 

It is clear that, in contrast to liquid metals, where the presence of the  
pseudopotential leads to a more complicated structure of the effective 
potential, in a dense hydrogen plasma, the effective potential is determined
only by  electron screening. 
As it was shown in \cite{Brovman} for liquid metals, the additional pair 
interaction, arising from third and fourth order terms in the expansion of 
the electron energy in terms of the pseudopotential can play an important role in the 
effective interaction. For the effective potential of a hydrogen plasma a
recent detailed analysis of these terms \cite{Kaim} showed that these terms 
are essential only for rather rarified plasma conditions ($r_s > 1.5$), and they are 
practically negligible for higher densities, $r_s < 1.5$, which we are considering 
in this paper. In fact, for  $r_s > 1.6$, the structure of the effective ion-ion
potential in hydrogen changes drastically and can be considered as precursor 
of the appearence of molecular states. In this paper, we will use the simplest
version of the method of an effective ion-ion potential (EIIP) which includes 
the electron-proton interaction up to second order, so we are restricted to 
sufficiently high densities, corresponding to $r_s<1.5$. 

Further progress can be made by using for $\Pi_e$ the 
random phase approximation (RPA), together with the long-wavelength and 
short-wavelength limits, 
 
\begin{eqnarray}\label{eq3.5}                                                             
\Pi^{\rm{RPA}}(q)&=& \Pi^{\rm{RPA}}(0) \left[1-\frac{1}{12} \frac{q^2}{q^2_F}\right]\,, 
\qquad q\ll q_F\,, 
\nonumber\\ 
\Pi^{\rm{RPA}}(q)&=& \Pi^{\rm{RPA}}(0) \frac{4}{3} \frac{q^2_F}{q^2}\,,  
\qquad q\gg q_F\,, 
\end{eqnarray}                                                                  
where $\hbar q_F=\sqrt{2m\epsilon_F}$  is the Fermi momentum of the electrons. 
The analysis of this expression shows that the main contribution to the 
energy (\ref{eq3.2}) comes from the small wave numbers.
Therefore, with sufficient accuracy, it is possible to neglect the q-dependence of $\Pi_e$ 
in Eq.~(\ref{eq3.2}) and, in particular, in the effective potential
(\ref{eq3.4}), replacing $\Pi^{\rm{RPA}}(q) \rightarrow \Pi^{\rm{RPA}}(0)$.
This means, we also neglect 
the well-known small oscillations of the effective potential for large 
distances, which are the result of a logarithmic singularity of the derivative 
$\left(d\Pi^{\rm{RPA}}/dq\right)|_{q=2q_F}$. For the densities under  
consideration (which are much higher than usual metallic densities), these oscillations 
are not essential for the thermodynamic functions. At the same time, 
it is crucial to calculate the polarization function $\Pi_e(0)$
fully selfconsistently: 
 
\begin{equation}
\label{eq3.6} 
\Pi_e(0)=\left(\frac{\partial n}{\partial \mu_e}\right)_T \,, \qquad  
\mu_e=\left(\frac{\partial {n \epsilon_e}}{\partial n}\right)_T, 
\end{equation} 
where $\epsilon_e$ is determined by (\ref{eq3.2}) and, consequently, 
takes into account the 
electron-electron exchange and correlations. For the case of degenerate 
electrons we can use one of the analytical approximations for $\epsilon_e$ such as, 
for example, that of Nozieres and Pines or Wigner, see e.g. \cite{mahan}
for an overview. 
Below we use Wigner's formula for the correlation energy, although for small
$r_s$ the approximation of Nozieres and  Pines is better (in fact, for
the region $r_s < 1$, where the deviations between these approximations for
the correlation energy become essential, we can neglect correlations at all
in comparison to kinetic and exchange terms). Because
$\Pi_e^{\rm{RPA}}(0)=\kappa^2_{TF}/(4\pi e^2)$   it is clear  that
Eq.~(\ref{eq3.6}) means renormalization of  $\Pi_e^{\rm{RPA}}\rightarrow
\Pi_e$   due to electron-electron  interaction and, therefore, a
renormalization of the momentum  $\kappa_{TF}\rightarrow\tilde{\kappa}_{TF}$:

\begin{eqnarray}\label{eq3.7}  
\Pi_e(0) &=& \Pi_e^{\rm{RPA}}(0) \,
\gamma(r_s)\,, \qquad \tilde{\kappa}_{TF}  \equiv \kappa_{TF}
\sqrt{\gamma(r_s)},  
\nonumber\\ 
\gamma(r_s) &=& 
\left(\frac{9\pi}{4}\right)^{2/3} \frac{6}{r_s^2} \frac{1} 
{r_s^2\frac{\partial^2\epsilon_e}{\partial r_s^2} -2r_s 
\frac{\partial \epsilon_e}{\partial r_s}}. 
\end{eqnarray} 
Because for the considered approximation the effective proton-proton 
potential is described by the screened potential of Thomas-Fermi type, see 
Eqs.~(\ref{eq3.4})-(\ref{eq3.7}): 
 
\begin{equation}\label{eq3.8} 
\Phi_{pp}(r)=\frac{e^2}{r} \,e^{-\frac{r}{\tilde{r}_{TF}}}, 
\end{equation} 
we conclude that there is renormalization of the screening radius which is 
due to electronic correlations: 
 
\begin{equation}\label{eq3.9} 
\tilde{r}_{TF} = \frac{1}{\tilde{\kappa}_{TF}} \equiv \frac{r_{TF}}{\sqrt{\gamma(r_s)}}. 
\end{equation} 
 
Let us now rewrite Eg.~(\ref{eq3.2}) for the considered approximation in the form: 
 
\begin{equation}\label{eq3.10} 
\epsilon=\epsilon_e + \epsilon_i
\end{equation}
\begin{equation}
\epsilon_i = \frac{3}{2} kT +\frac{1}{2N} \sum_{i\neq j} \Phi_{pp} (R_i-R_j) - 
\frac{e^2}{d} \left(\frac{\kappa}{2}+ \frac{3}{2\kappa^2}\right), 
\end{equation} 
where $\kappa\equiv d\cdot\tilde{\kappa}_{TF}$. 
After averaging over the proton positions with a Gibbs distribution
(denoted by $\langle \dots \rangle$), 
Eq. (\ref{eq3.10}) can be represented as the sum of two terms: 
$\epsilon_e$ - the energy of a degenerate electron liquid 
on the positive homogeneous charge background  and the energy of screened 
classical charged protons, interacting via the screened potential 
(\ref{eq3.9}) and renormalized by the constant terms, obtained above: 

\begin{equation}\label{eq3.11}
\epsilon_i = \left(u+\frac{3}{2}\right) kT,
\end{equation}
with
\begin{equation}\label{eq3.12} 
\quad u \equiv \Gamma \left\{ \frac{d}{2Ne^2} 
\left\langle\sum_{i\neq j} \Phi_{pp}(R_i-R_j)\right\rangle - \frac{\kappa}{2} - 
\frac{3}{2\kappa^2}\right\}. 
\end{equation}                                                                 
Here $u$ is the ionic interaction energy in $kT-$ units.
The energy (\ref{eq3.11}) coincides with accuracy $(kT/E_F)^2$ with the usual 
thermodynamic energy determined from the free energy of the system because, 
in the considered parameter range, the electrons are degenerate (with the 
same accuracy).
From expression (\ref{eq3.11}) follows that the energy of a classical 
one-component system of charged particles interacting via a screened 
(Debye or Yukawa) potential tends to infinity as 
$3k_BT\Gamma/2\kappa^2$ for $\kappa\rightarrow 0$ 
(i.e. the screening radius diverges). 
The function $u/\Gamma$ has been tabulated in \cite{Hamaguchi,Farouki} 
(for the calculations of the phase diagram of a purely classical 
one-component Debye plasma), as function of the two parameters $\Gamma$ and
the dimensionless screening length $\kappa$, based on accurate MD
calculations for the Debye system. Below we 
use these numerical results to calculate the energy of a dense
hydrogen plasma 
in the described above approximations. 
Within the Wigner approximation for the electron energy,  
                                                                              
\begin{eqnarray}\label{eq3.13} 
\epsilon_e&=& \left( \frac{2.21}{r_s^2}- \frac{0.916}{r_s}+ \epsilon_{\rm{corr}}\right) Ry, 
\nonumber\\ 
\epsilon_{\rm{corr}}&=& - \frac{0.88}{r_s+7.8},
\end{eqnarray} 
we obtain, from Eq.~(\ref{eq3.7}): 
 
\begin{eqnarray}\label {eq3.14}
\gamma(r_s)&=& \frac{22.1}{r_s^2\varphi(r_s)},
\nonumber\\ 
\varphi(r_s)&=& \frac{22.1}{r_s^2}- \frac{3.664}{r_s} - \frac{1.76 r_s}{(r_s+7.8)^2}  
- \frac{1.76 r_s^2}{(r_s+7.8)^3},
\end{eqnarray}      
where $\gamma(r_s\rightarrow 0) \rightarrow 1$. 
Now, the total internal energy, Eq.~(\ref{eq3.11}),  can be expressed 
in terms of the tabulated function $u/\Gamma$ as:  
 
\begin{equation}\label{eq3.15} 
\epsilon=\left[\frac{2.21}{r_s^2}- \frac{0.916}{r_s} + \epsilon_{\rm{corr}} + \frac{2}{r_s} 
\left(\frac{u}{\Gamma}+ \frac{3}{2\Gamma}\right)\right] Ry. 
\end{equation} 
The numerical results computed from this approximation are included in Figs. 
\ref{et10}--\ref{et50} below.   
 
Alternatively, we may use additional approximations for the computation of the internal 
energy of the plasma. This can be done by averaging Eq.~(\ref{eq3.2}) over the ion 
Gibbs distribution with the same effective 
Hamiltonian (\ref{eq3.2}). Than we immediately find for the average energy per 
proton,
 
\begin{eqnarray}\label{eq3.16} 
\frac{\langle E\{R_i\}\rangle}{N_p} &=& \epsilon_e + \frac{3}{2} k_BT - \frac{1}{2} \int \frac{d^3q}{(2\pi)^3} 
\frac{u_{ei}^2(q)\Pi_e(q)}{\varepsilon_e(q)}+\nonumber\\ 
&+& \frac{1}{2}\int \frac{d^3q}{(2\pi)^3} {\cal V}_{ii}^{\rm{eff}}(q) 
\left[S_{ii}(q)-1\right]= 
\nonumber\\ 
&=& \epsilon_e+ \frac{3}{2} k_BT + \frac{1}{2} \int \frac{d^3q}{(2\pi)^3} u_{ii}(q) 
\left[S_{ii}(q)-1\right] -\frac{1}{2} 
\int \frac{d^3q}{(2\pi)^3} \frac{u_{ei}^2\Pi_e(q)}{\varepsilon_e (q)} 
S_{ii}(q), 
\end{eqnarray} 
where we introduced the ion-ion structure factor 
$S_{ii}(k)$ defined as
 
\begin{eqnarray}\label{eq3.17} 
\langle \varrho_{\vec{k}_1}\varrho_{\vec{k}_2}\rangle &=& 
N S_{ii}(\vec{k}_1) \delta_{\vec{k}_1+\vec{k}_2,0} 
+ N^2 \delta_{\vec{k}_1,0} \delta_{\vec{k}_2,0}\nonumber\\ 
\varrho_k &\equiv&  \sum_j e^{-i\vec{k}\vec{R}_j}\,, \qquad \delta_{k,0}= \left\{ 
\begin{array}{ll} 
1, & k=0\\ 
0, & k\neq 0
\end{array}\right. 
\end{eqnarray}

Eq.~(\ref{eq3.16}) can be simplified by replacing, approximately, the full 
structure factor by the OCP structure factor $S_{ii}^{\rm{OCP}}$, computed with 
the effective ion--ion interaction. Then, the full energy can be written as the sum 
of three contributions: the first from the electron subsystem, the second from the 
classical ion OCP subsystem (both imbedded, respectively, into a positive and 
negative charge background) and a third term, $\epsilon_i^{POL}$, which
describes in perturbation-theoretical approximation for the polarization of the
electron liquid by the ions. The resulting formulas coincide 
with the perturbation approximations derived by Hansen, DeWitt and others  
\cite{Hansen,DeWitt}:
   
\begin{eqnarray}\label{eq3.19a} 
\frac{\langle E\{R_i\}\rangle}{N_p} &=& \epsilon_e + \epsilon_i^{\rm{OCP}}+\delta \epsilon\,, 
\\ 
\label{eq3.20a} 
\delta \epsilon &=& \frac{e^2}{\pi} \int\limits_0^\infty dq 
\left(\frac{1}{\varepsilon_e(q)} -1\right) 
S_{ii}^{\rm{OCP}}(q). 
\end{eqnarray}  
As is clear from the above derivations,  Eqs.~(\ref{eq3.19a}), (\ref{eq3.20a}) 
are less accurate than the full EIIP model presented above.      
   
\section{Pad\'e approximations and chemical picture: PACH method}\label{pach}

In this section we will explain in brief the method of Pad\'e approximations
in combination with the chemical picture, i.e. Saha equations
\cite{green-book,Pade85,Pade90,EbFo91} (PACH). 
On the basis of the PACH-approximation we
will calculate the internal energy for the 3 isotherms
$T = 10,000$K, $30,000$K, and $50,000$K
for those regions of the
density where bound states (atoms and molecules) play a minor role.
In other words we restrict our study to the density region where 
the plasma is strongly (but not necessarily fully) ionized.
This method works only with analytical formulae which
are, however, rather complicated; nevertheless the calculation of
one energy data point takes no more
than a few seconds on a PC.

The Pad\'e approximations were constructed in earlier work from the known analytical results
for limiting cases of low density \cite{green-book,ebeling} and
high density \cite{green-book}. The structure of the Pad\'e approximations was 
devised in such a way that
they are analytically exact up to
quadratic terms in the density (up to the second virial coefficient) and interpolate between
the virial expansions and the high-density asymptotic expressions
\cite{Pade85,Pade90,EbFo91}. The formation of bound states was taken into account
by using a chemical picture.

We follow in large here this cited work, only the contribution
of the OCP-ion-ion interaction which is, in most cases, the largest one, was
substantially improved following \cite{Kahlbaum96}. With respect to the
chemical picture we restricted ourselves to the region of strong ionization
where the number of atoms is still relatively low and where no molecules are
present. We will discuss here only the general structure of the Pad\'e
formulae.  The internal energy density of the plasma is given by

\begin{equation}
E = E_{id} + E_{int}
\label{eq4.1}
\end{equation}
Here $E_{id}$ is the internal energy of an ideal plasma consisting of Fermi
electrons, classical protons and classical atoms and $E_{int}$ is the
interaction energy which is represented by

\begin{equation}
E_{int} = N_p \left(\epsilon_{e} + \epsilon_{i} + \epsilon_a \right)
\label{eq4.2}
\end{equation}

The splitting of the interaction contribution to the internal energy 
corresponds largely to the previous section. We have:
\begin{itemize}
\item The electron-electron interaction: This term corresponds to the OCP
energy  of the electron subsystem. Instead of the simple expressions 
used in earlier work \cite{Pade85,Pade90,FiBoEbFo01} we used here a more refined
formula for the energy \cite{EKKR86}. This formula is an interpolation between
the Hartree limit with the Gellman-Brueckner correction
(used already in the previous section), the Wigner limit and the Debye law
including quantum corrections\\

\begin{equation}
\epsilon_{e} = -\frac{(r_s^3 + 50)\left[a_H+a_W(r_s)\right] +2\sqrt{6} \,d_0\, r_s^{5.5}\, \tau^{2.5}
+24 \,d_H \,r_s^4 \, \tau^2}
{(r_s^3+50)r_s +2.3\, r_s^4 \, \tau^2 + 2\sqrt{6} \, d_1 \, r_s^{5.5} \, \tau^2
+ r_s^7 \, \tau^3}. 
\label{eq4.3}
\end{equation}
Here, a Wigner function has been introduced which is given by
\begin{equation}
a_W(x) = 2 \,b_0\, x \log \left(1+\left[x^{0.5} e^{-b_1/(2b_0)} +2 b_0 x /a_W\right]^{-1}\right),
\label{eq4.4}
\end{equation}
and the constants have the values $d_0 = 0.5; d_1=0.6631; d_H=0.125; a_H=0.91633; a_W=0.87553; 
b_0=0.06218;$ and $b_1=0.0933$. We mention that similar formulas are valid also 
for other thermodynamic
functions by adjusting the constants \cite{EKKR86}.
The formula we have used here for the OCP contains all terms taken into
account in the previous section but, in addition, also temperature
dependent corrections.

\item The ion contribution to the internal energy $\epsilon_i$.
This term was calculated in the previous section. Here we will use
a procedure which is based on the approximation 
(\ref{eq3.17}, \ref{eq3.19a}). This enables us
to use results of the MC-calculations of Hansen, DeWitt and others
\cite{DeWitt,Slattery80}.
According to Eqs.~(\ref{eq3.17}, \ref{eq3.19a}) the ion contribution is 
split into two terms

\begin{equation}
\epsilon_i = \epsilon_{i}^{OCP} + \epsilon_{i}^{POL},
\label{eq4.5}
\end{equation}
where the first represents the OCP-contribution of the protons
and the second the polarization of the proton OCP by the electron
gas. For the region of high densities, i.e. large $\Gamma$ and small
$r_s$ we use the Livermore Monte Carlo data  which were
parametrized by DeWitt in the form

\begin{eqnarray}\label{eq4.6}
\epsilon_{i}^{OCP} &=& - .8946 \Gamma  + .8165 \Gamma^{.25} -.5012,
\\
\epsilon_{i}^{POL} &=& - r_s \,(.0543 \Gamma  + .1853 \Gamma^{.25} -.0659).
\label{eq4.7}
\end{eqnarray}
We note that the polarization term describes the correction due to screening
of the proton-proton interaction by the electron fluid. 
In order to obtain these expressions,
semiclassical MC calculations were performed based on effective ion interactions
which model the electrons as a responding background \cite{Hansen,DeWitt}. We
do not  need to go into the details of this method since the procedure
corresponds to Eq.~(\ref{eq3.20a}) derived in the last section.\\
In the low density limit we used the Debye law with quantum corrections
\cite{green-book,EbFo91}

\begin{eqnarray}
\label{eq4.8}
\epsilon_{i}^{OCP} &=& - .86603 \tau d_0 \Gamma^{1.5} [1 - B_1 \Gamma^{1.5}],
\\
\epsilon_{i}^{POL} &=& - .71744 \Gamma^{1.5} [1 - C_1 \Gamma^{1.5}].
\label{eq4.9}
\end{eqnarray}
Here, the temperature functions $B_1$ and $C_1$ describe rather complex
quantum corrections which 
are, however, explicitly known and are easily programmed \cite{green-book}.
The Pad\'e approximations which connect the high and the low density limits
are constructed by standard methods \cite{Pade85,Pade90,EbFo91} and will not be
given here in  explicit form. For the OCP-energy of the ions we use
the very accurate formulas  proposed by Kahlbaum \cite{Kahlbaum96}.

\item The atomic contribution:
In the region of
densities and temperatures which is studied in this work this
contribution gives only a small correction. We calculate the number
of atoms on the basis of a nonideal Saha equation. The formation of
molecules is not taken into account. 
We restrict the calculations to a region where the number
density of atoms is so small, that the degree of ionization 
is larger than $75 \%$. 
The contributions to the chemical potential which appear in
the Saha equation are calculated, in part,
from scaling relations and, in part by numerical
differentiation of the free energy given earlier \cite{Pade85,Pade90}.
For the partition function in the Saha equation we use the Brillouin-Planck-Larkin
expression \cite{green-book,EbFo91}. The nonideal Saha
equation which determines the degree of ionization (the density of the atoms)
is solved by 5-100 iterations starting  from the ideal Saha equation.
Due to the high degree of ionization, the atomic interaction contributions 
can be approximated in the simplest way by the second virial contribution and
by treating the atoms as small hard spheres.

\end{itemize}
The results of our Pad\'e calculations for a broad density interval for 
three isotherms are included in Figs.~\ref{et10}--\ref{et50}.

\section{Summary of the Path integral Monte Carlo simulations}\label{pimc}

The analytical approximations discussed in the previous sections 
work very well at high densities and if bound states are of minor 
importance. These conditions are not fulfilled for densities below 
than the Mott point corresponding to $r_s>1$. Here, recently 
developed path integral Monte Carlo simulations can be used.
Starting from the basic plasma particles, electrons and ions, 
they ``automatically'' account for bound state formation and 
ionization and dissociation. Furthermore, in contrast to a 
chemical picture, no restrictions on the type of chemical 
species are made and the appearance of complex aggregates such 
as molecular ions or clusters of several atoms are fully 
included. On the other hand, the simulations are becoming 
increasingly difficult at high density where the electron 
degenercy is large.
For this reason it is of high interest to compare results 
of the PIMC approach with alternative theories as they are 
expected to complement each other. This will be done 
in the next section.

But first, we briefly outline the idea of our direct PIMC scheme.
All thermodynamic properties of a two-component plasma are defined
by the partition function $Z$ which, for the case of $N_e$
electrons and $N_p$ protons, is given by
\begin{eqnarray}
Z(N_e,N_p,V,\beta) &=&
\frac{Q(N_e,N_p,\beta)}{N_e!N_p!},
\nonumber\\
\mbox{with} \qquad
Q(N_e,N_p,\beta) &=& \sum_{\sigma}\int\limits_V dq \,dr
\,\rho(q, r,\sigma;\beta),
\label{q-def}
\end{eqnarray}
where $\beta=1/k_B T$.
The exact density matrix is, for a quantum system, in
general, not known but can be constructed using a path integral representation
\cite{feynman-hibbs},
\begin{eqnarray}
\int\limits_{V} dR^{(0)}\sum_{\sigma}\rho(R^{(0)},\sigma;\beta) &=&
 \int\limits_{V} dR^{(0)} \dots dR^{(n)} \,
\rho^{(1)}\cdot\rho^{(2)} \, \dots \rho^{(n)}
\nonumber\\
&\times&\sum_{\sigma}\sum_{P} (\pm 1)^{\kappa_P}
\,{\cal S}(\sigma, {\hat P} \sigma')\,
{\hat P} \rho^{(n+1)},
\label{rho-pimc}
\end{eqnarray}
where $\rho^{(i)}\equiv \rho\left(R^{(i-1)},R^{(i)};\Delta\beta\right) \equiv
\langle R^{(i-1)}|e^{-\Delta \beta {\hat H}}|R^{(i)}\rangle$,
whereas $\Delta \beta \equiv \beta/(n+1)$ and
$\Delta\lambda_a^2=2\pi\hbar^2 \Delta\beta/m_a$, $a=p,e$.
${\hat H}$ is the Hamilton operator, ${\hat H}={\hat K}+{\hat U}_c$, containing
kinetic and potential energy contributions, ${\hat K}$ and ${\hat U}_c$,
respectively, with
${\hat  U}_c = {\hat  U}_c^p + {\hat  U}_c^e + {\hat  U}_c^{ep}$ being the sum
of the Coulomb potentials between protons (p), electrons (e) and electrons and
protons (ep). Further,
$R^{(i)}=(q^{(i)},r^{(i)}) \equiv (R_p^{(i)},R_e^{(i)})$, for $i=1,\dots n+1$,
$R^{(0)}\equiv (q,r)\equiv (R_p^{(0)},R_e^{(0)})$, and
$R^{(n+1)} \equiv R^{(0)}$ and $\sigma'=\sigma$. This means, the particles are
represented by fermionic loops with the coordinates (beads)
$[R]\equiv [R^{(0)}; R^{(1)};\dots; R^{(n)}; R^{(n+1)}]$, where $q$ and
$r$ denote the electron and proton coordinates, respectively.
The spin gives rise to the spin part of the density matrix ${\cal S}$, whereas
exchange effects are accounted for by the permutation operator ${\hat P}$, 
which acts on the electron coordinates and spin projections, and
the sum over the permutations with parity $\kappa_P$. In the fermionic case
(minus sign), the sum contains $N_e!/2$ positive and negative terms leading
to the notorious sign problem. Due to the large mass difference of electrons
and ions, the exchange of the latter is not included.

To compute thermodynamic functions, the logarithm of the partition function
has to be differentiated with respect to thermodynamic variables. 
In particular,  the internal energy $E$ follows from $Q$ by 
\begin{eqnarray}
\beta E &=& -\beta \partial {\rm ln} Q
/ \partial \beta,
\label{e_gen}
\end{eqnarray}
This leads to the following result (for details, cf. \cite{FiBoEbFo01}),
\begin{eqnarray}
\beta E = \frac{3}{2}(N_e+N_p) + \frac{1}{Q}
\frac{1}{\,\lambda_p^{3N_p}\Delta \lambda_e^{3N_e}}\sum_{s=0}^{N_e}
\int dq \, dr \, d\xi \,\rho_s(q,[r],\beta) \,\times
\nonumber\\
\Bigg\{\sum_{p<t}^{N_p} \frac{\beta e^2}{|q_{pt}|} +
\sum_{l=0}^{n}\Bigg[\sum_{p<t}^{N_e} \frac{\Delta\beta e^2}{|r^l_{pt}|}
+  \sum_{p=1}^{N_p}\sum_{t=1}^{N_e} \Psi_l^{ep}\Bigg]
\nonumber\\
+ \sum_{l=1}^{n}\Bigg[
- \sum_{p<t}^{N_e}C^l_{pt}
\frac{\Delta\beta e^2}{|r^l_{pt}|^2} +
 \sum_{p=1}^{N_p}\sum_{t=1}^{N_e}
D^l_{pt}
\frac{\partial \Delta\beta\Phi^{ep}}{\partial |x^l_{pt}|}
 \Bigg]
\nonumber\\
\,-\,
\frac{1}{{\rm det} |\psi^{n,1}_{ab}|_s}
\frac{\partial{\rm \,det} | \psi^{n,1}_{ab} |_s}{\partial \beta}
\Bigg\},
\nonumber \\
{\rm with} \quad C^l_{pt} = \frac{\langle r^l_{pt}|y^l_{pt}\rangle}{2|r^l_{pt}|},
\qquad D^l_{pt} = \frac{\langle x^l_{pt}|y^l_{p}\rangle}{2|x^l_{pt}|},
\label{energy}
\quad
\end{eqnarray}
and $\Psi_l^{ep}\equiv \Delta\beta\partial
[\beta'\Phi^{ep}(|x^l_{pt}|,\beta')]/\partial\beta'|_{\beta'=\Delta\beta}$
contains the electron-proton Kelbg potential $\Phi^{ep}$, 
cf. Eq.~(\ref{kelbg}) below.
Here,
$\langle \dots | \dots \rangle$ denotes the scalar product, and
$q_{pt}$, $r_{pt}$ and $x_{pt}$ are differences of two
coordinate vectors:
$q_{pt}\equiv q_p-q_t$,
$r_{pt}\equiv r_{p}-r_{t}$, $x_{pt}\equiv r_p-q_t$, $r^l_{pt}=r_{pt}+y_{pt}^l$,
 $x^l_{pt}\equiv x_{pt}+y^l_p$ and
 $y^l_{pt}\equiv y^l_{p}-y^l_{t}$, with $y_a^n=\Delta\lambda_e\sum_{k=1}^{n}\xi^{(k)}_a$.
Here we introduced dimensionless distances between neighboring
vertices on the loop, $\xi^{(1)}, \dots \xi^{(n)}$,
thus, explicitly,
$[r]\equiv [r; y_e^{(1)}; y_e^{(2)}; \dots].$ Further,
the density matrix $\rho_s$ in Eq. (\ref{energy}) is given by
\begin{eqnarray}
\rho_s(q,[r],\beta) = C^s_{N_e}
\, e^{-\beta U(q,[r],\beta)} \prod\limits_{l=1}^n
\prod\limits_{p=1}^{N_e} \phi^l_{pp}
{\rm det} \,|\psi^{n,1}_{ab}|_s,
\label{rho_s}
\end{eqnarray}
where
$U(q,[r],\beta)=
U_c^p(q)+\{U^e([r],\Delta\beta)+U^{ep}(q,[r],\Delta\beta)\}/(n+1)$
and  $\phi^l_{pp}\equiv \exp[-\pi |\xi^{(l)}_p|^2]$.
We underline that the density matrix (\ref{rho_s})
does not contain an explicit
sum over the permutations and thus no sum of terms with alternating
sign. Instead, the whole exchange problem is
contained in a single exchange matrix given by
\begin{eqnarray}
||\psi^{n,1}_{ab}||_s\equiv ||e^{-\frac{\pi}{\Delta\lambda_e^2}
\left|(r_a-r_b)+ y_a^n\right|^2}||_s.
\label{psi}
\end{eqnarray}
As a result of the spin summation,
the matrix carries a subscript $s$ denoting the number of electrons having
the same spin projection.

The potential $\Phi^{ab}$ appearing in Eq.~(\ref{energy}) is an 
effective quantum pair interaction between two charged particles
immersed into a weakly degenerate plasma. It has been derived by 
Kelbg and co-workers \cite{Ke63,kelbg} who showed that it contains 
quantum effects exactly in first order in the coupling parameter $\Gamma$,
\begin{eqnarray}
\Phi^{ab}(|{\bf r}_{ab}|,\Delta\beta) =
\frac{e_a e_b}{\lambda_{ab} x_{ab}} \,
\left\{1-e^{-x_{ab}^2}+\sqrt{\pi}\, x_{ab}
\left[1-{\rm erf}(x_{ab})\right]
\right\},
\label{kelbg}
\end{eqnarray}
where $x_{ab}=|{\bf r}_{ab}|/\lambda_{ab}$, and we underline that the Kelbg
potential is finite at zero distance.

The structure of Eq.~(\ref{energy}) is obvious: 
we have separated the classical
ideal gas part (first term). The ideal quantum part in excess of the
classical one and the correlation
contributions are contained in the
integral term, where the second line results from the ionic correlations
(first term) and the e-e and e-i interaction at the first vertex (second
and third terms respectively). 
Thus, Eq.~(\ref{energy}) contains the important
limit of an ideal quantum plasma in a natural way.
The third and fourth lines are due to the
further electronic vertices and the explicit
temperature dependence [in Eq.~(\ref{energy})] and volume dependence (in
the corresponding equation of state result) of the exchange matrix, 
respectively.
The main advantage of Eq.~(\ref{energy}) is that the
explicit sum over permutations has been converted into the spin determinant
which can be computed very efficiently using standard linear algebra
methods. Furthermore, each of the sums in curly brackets in
Eq.~(\ref{energy}) is bounded as the number of vertices increases,
$n\rightarrow \infty$. The error of the total expression is of the
order of $1/n$. Thus, expression (\ref{energy}) and the analogous
result for the equation of state are well suited for numerical
evaluation using standard Monte Carlo techniques, e.g. 
\cite{Zamalin,binder96}.
 
In our Monte Carlo scheme
we used three types of steps, where either electron or proton coordinates,
$r_i$ or $q_i$  or
inidividual electronic beads $\xi_i^{(k)}$ were moved until convergence of the
calculated values was reached. Our procedure has been extensively tested.
In particular, we found from comparison with the known analytical expressions
for pressure and energy of an ideal Fermi gas that the Fermi statistics is
very well reproduced \cite{filinov-etal.00jetpl}. Further, we performed
extensive tests for few--electron systems in a harmonic trap where, again,
the analytically known limiting behavior (e.g. energies) is well reproduced
\cite{afilinov-etal.00pss,afilinov-etal.99prl}.
For the present simulations of dense hydrogen, we varied both the particle
number and the number of time slices (beads). As a result of these tests,
we found that to obtain convergent results for the thermodynamic properties
of hydrogen in the density-temperature region of interest here, 
particle numbers $N_e=N_p= 50$ and beads numbers in
the range of $n=6\dots 20$ are an acceptable compromise between 
accuracy and computational effort
\cite{filinov-etal.99xxx1,filinov-etal.00jetpl,filinov-etal.00pla}.

\section{Numerical Results. Comparison of the analytical and simulation data}
\label{res}

\begin{figure}[p]
\centerline{
\psfig{file=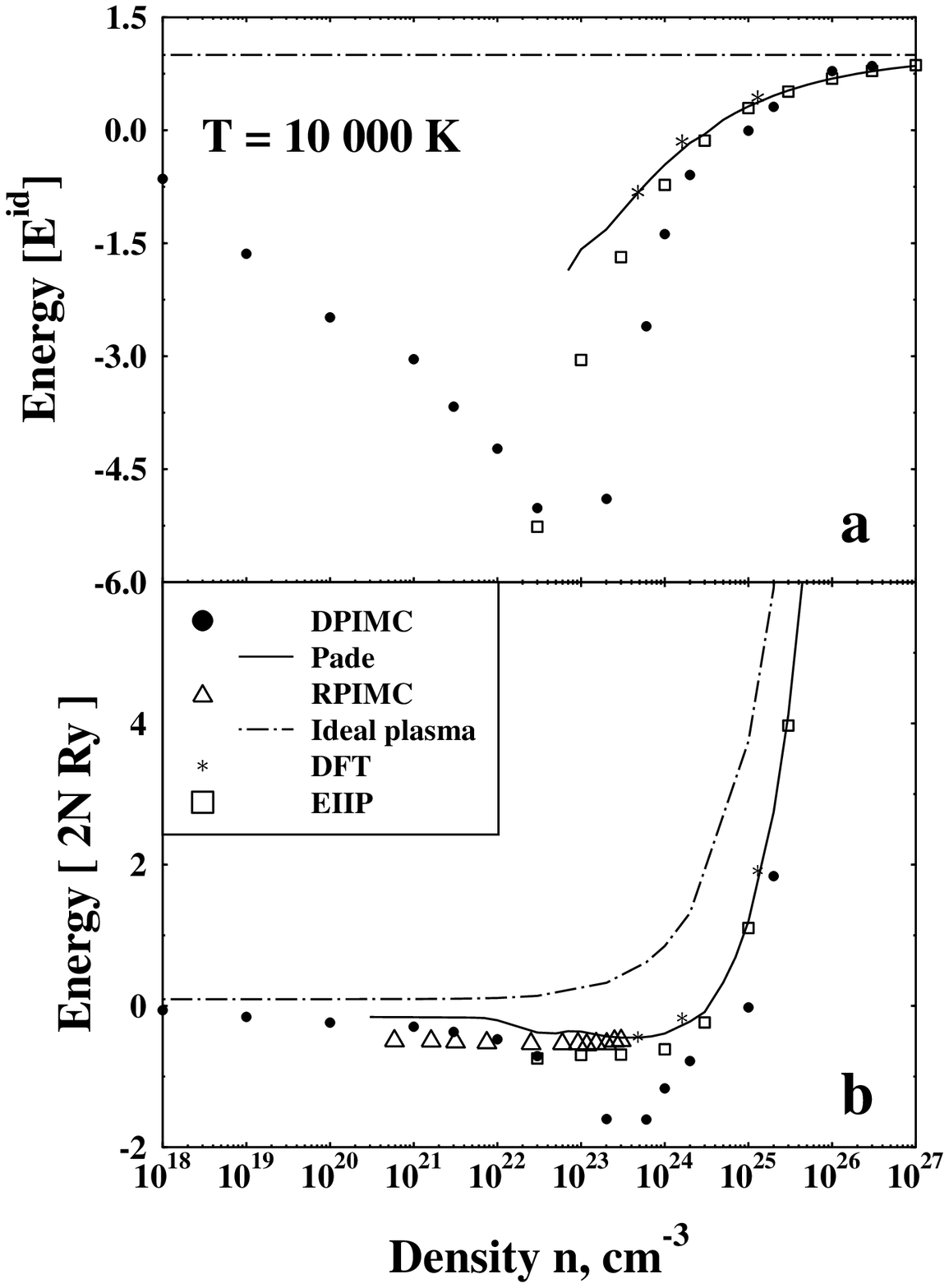,height=18cm}}
\caption[]{\label{et10}
Internal Energy of hydrogen for $T=10,000K$, a) normalized
to the energy of a noninteracting electron-proton system and
b) in units of 2N Rydberg. The curves show results of
PACH-calculations (``Pade''), the EIIP model, our Monte Carlo
simulations (``DPIMC''), density functional theory (``DFT'') \cite{Xu98}
and restricted PIMC data (``RPIMC'') of Militzer et al. \cite{militzer-etal.00}.
}
\end{figure}

\begin{figure}[p]
\centerline{
\psfig{file=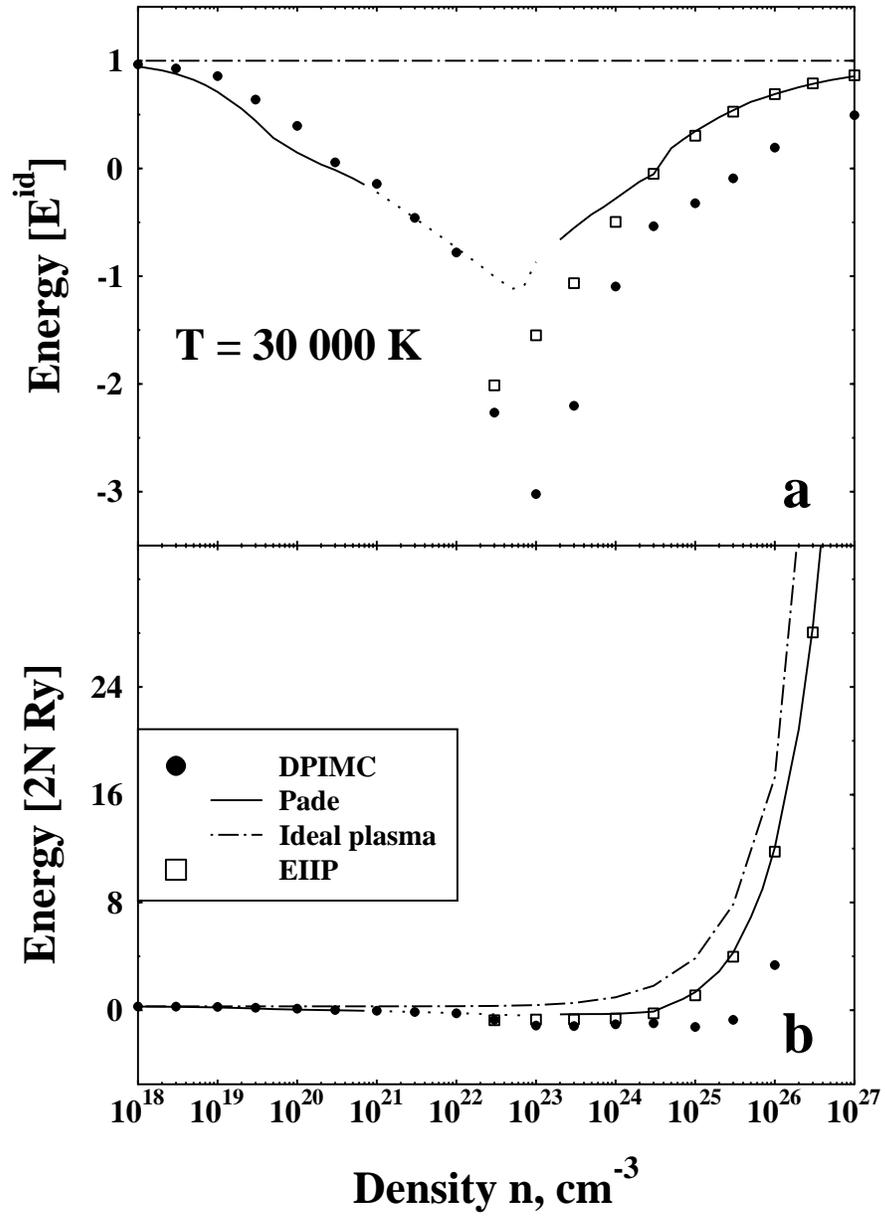,height=18cm}}
\vspace{-1cm}
\caption[]{\label{et30}
Internal Energy of hydrogen for $T=30,000K$. Same notation as in
Fig.~\ref{et10}. Dotted line indicates region of low degree of ionization where
the Pade and EEIP results are less reliable.
}
\end{figure}

\begin{figure}[p]
\centerline{
\psfig{file=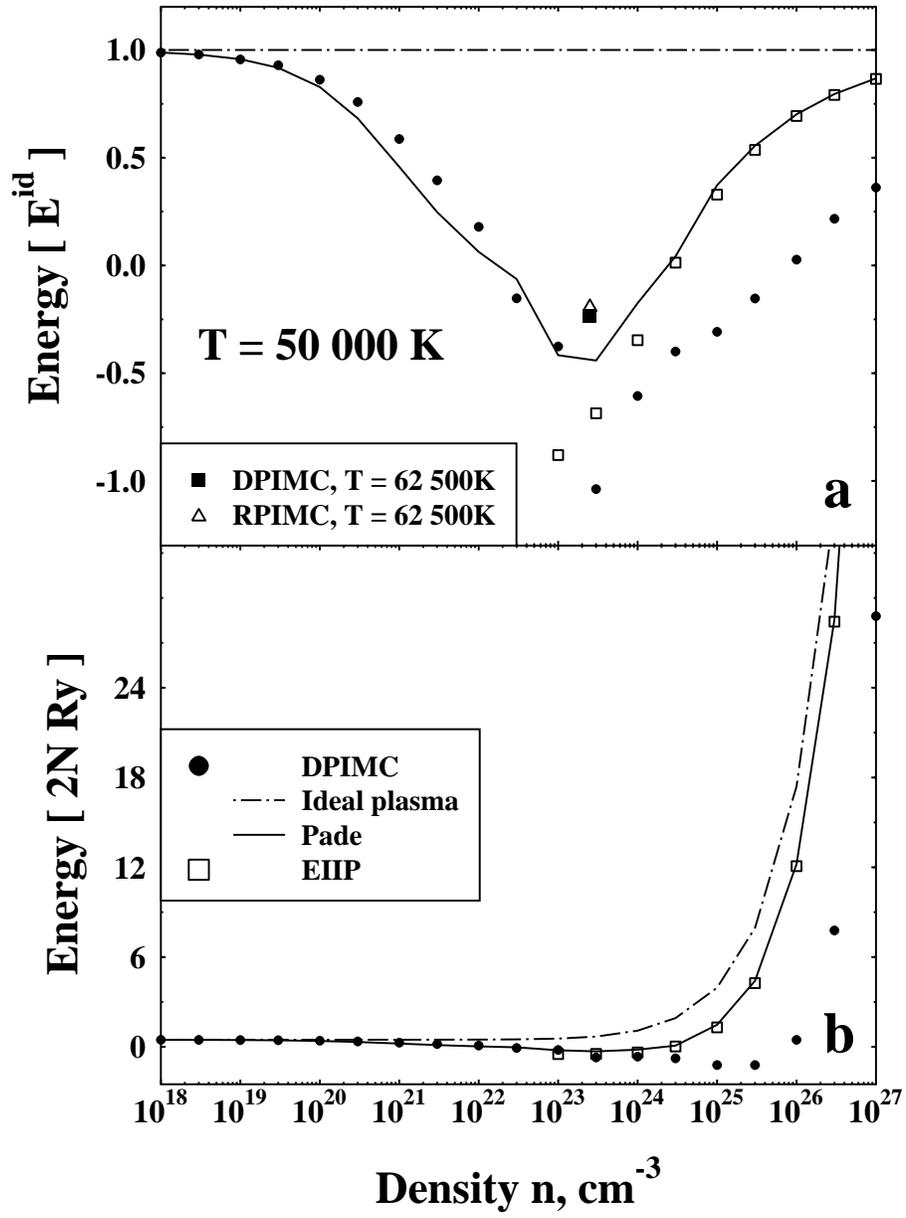,height=18cm}}
\vspace{-1cm}
\caption[]{\label{et50}
Internal Energy of hydrogen for $T=50,000K$. Same notation as in
Fig.~\ref{et10}.
}
\end{figure}

Let us now  come to the numerical results. We have computed the internal 
energy of dense hydrogen using the two analytical (EIIP and PACH) approaches 
and the PIMC simulations. The data are shown in Figs.~\ref{et10}--\ref{et50} 
for three temperatures, $10,000$K, $30,000$K and $50,000$K, respectively.

Consider first the general behavior which is clearly seen for the lowest 
temperature, cf. Fig.~\ref{et10}.b. The overall trend is an increase 
of the energy with density which is particularly rapid at high densities 
due to electron degeneracy effects; this is clearly seen from the 
{\em ideal plasma} curve (dash-dotted line). The {\em nonideal plasma} results
show a prominent deviation from this trend which is in full agreement 
with the discussion given in Section \ref{par}: the formation of an energy 
minimum (where the energy may become negative) at intermediate densities.
Our calculations for a nonideal 
hydrogen plasma asymptotically approach the ideal curve both, at low density 
(ideal classical plasma) and at high density (ideal mixture of classical 
protons and quantum electrons).
At intermediate densities, between $10^{19}$cm$^{-3}$ and $10^{25}$cm$^{-3}$,
the nonideal plasma energy is significantly
lower than the ideal energy which is due to strong
correlations and formation of bound states. In particular, we see clearly
that indeed, for the considered temperatures, the total energy reaches
negative values.

Let us now compare the results from the different methods. First, we see that
the energy minimum is reproduced by
all methods, but there are quantitative differences regarding its
depth and width. The general observation made for all temperatures,
cf. also Figs.~\ref{et30} and \ref{et50},
is that the simulations yield a  deeper minimum and
shift of the energy increase towards higher densities.
Before further
analyzing these differences, we concentrate on the results of the
analytical approaches. For all temperatures (including higher ones),
the PACH and EIIP approaches  coincide in the limit of
high densities. This is an important test since both contain the
ideal Fermi gas result as a limiting case for high degeneracy. The
interesting result is that this agreement holds up to densities
as low as $n=10^{24}$cm$^{-3}$. For still lower densities, the EIIP method
yields lower energies which are closer to the PIMC results. At these
densities, atom and molecule formation is becoming important, and both
analytical methods (in their present form) are becoming unreliable.
(For this reason, the Pad\'e curve in Fig.~\ref{et10}.a is discontinued
below $10^{23}$cm$^{-3}$, and in Fig.~\ref{et30} the uncertain region is
indicated by the dotted curve.)

It interesting compare to another theoretical approach based on density
functional (DFT) calculations. Recently, Xu and Hansen \cite{Xu98} published data
for $T=10,000$K and $r_s\le 1.5$ which are also included in Fig.~\ref{et10}.
These calculations which also neglect bound state formation practically
coincide with the PACH results. The good agreement of the three
completely independent approaches - EIIP, PACH and DFT - is a strong indication that
they are able to yield reliable results for a  fully ionized macroscopic hydrogen plasma
at high densities, $r_s\le 1.5$.

Let us now turn to the comparison with the PIMC simulations. As noted above,
the overall agreement of all methods is satisfactory in view of the
strength of correlation and quantum effects. Nevertheless, we observe deviations
of our PIMC results from all other data, in particular around the energy minimum.
Our data for $T=10,000$K are also lower than restricted PIMC results of
Militzer et al. \cite{militzer-etal.00}, cf. Fig.~\ref{et10}.b,
whereas we found excellent quantitative agreement between the two independent quantum
Monte Carlo methods above $T=50,000$K, see the point for $T=62,500$K in
Fig.~\ref{et50}.a, (see also Ref. \cite{FiBoEbFo01}). The reason for
the low energies observed in our PIMC simulations at $T=10,000$K are finite size
effects: the homogeneous plasma state is unstable in the density region of the
energy minimum. An analysis of the electron-proton configurations reveals that the
plasma gains energy by forming small droplets \cite{filinov-etal.01jetpl} which is
a direct indication for a first order phase transition as discussed in the Introduction.
These effects begin to appear in the weakly ionized plasma and are not contained
in the present variants of the PACH and EIIP methods although they have been
analyzed before \cite{ppt}. It is interesting to
note that Xu and Hansen \cite{Xu98} 
observed strong fluctuations in their density functional calculations below $r_s=1.5$ 
which strongly resembled precursors of a phase transition. To clearify this interesting 
issue more in detail requires extensive simulations which are presently under way. 

For completeness, we mention further effects which tend to lower the
total energy and which are neglected in the analytical approaches: increased 
electron polarization and nonadditive terms in the efficient proton-proton
interaction which were analyzed by Kagan and co-workers  \cite{Brovman}.

The next interesting feature of the PIMC simulations is the shift of the
energy growth to higher density values compared to the analytical models. This tendency
becomes stronger with increasing temperature, as can be seen in
Figs.~\ref{et10}--\ref{et50}. There is no reason to doubt that the analytical methods 
(in accord with the density functional results) yield the correct energy asymptotics 
of a macroscopic electron-proton plasma at very high densities. (Account of proton degeneracy
effects which are not included would only further increase the energies.) As noted above,
our direct PIMC simulations become increasingly difficult with growing electron degeneracy, 
so we expect the results to become less accurate for densities exceeding $10^{25}$cm$^{-3}$.

\begin{figure}[p]
\centerline{
\psfig{file=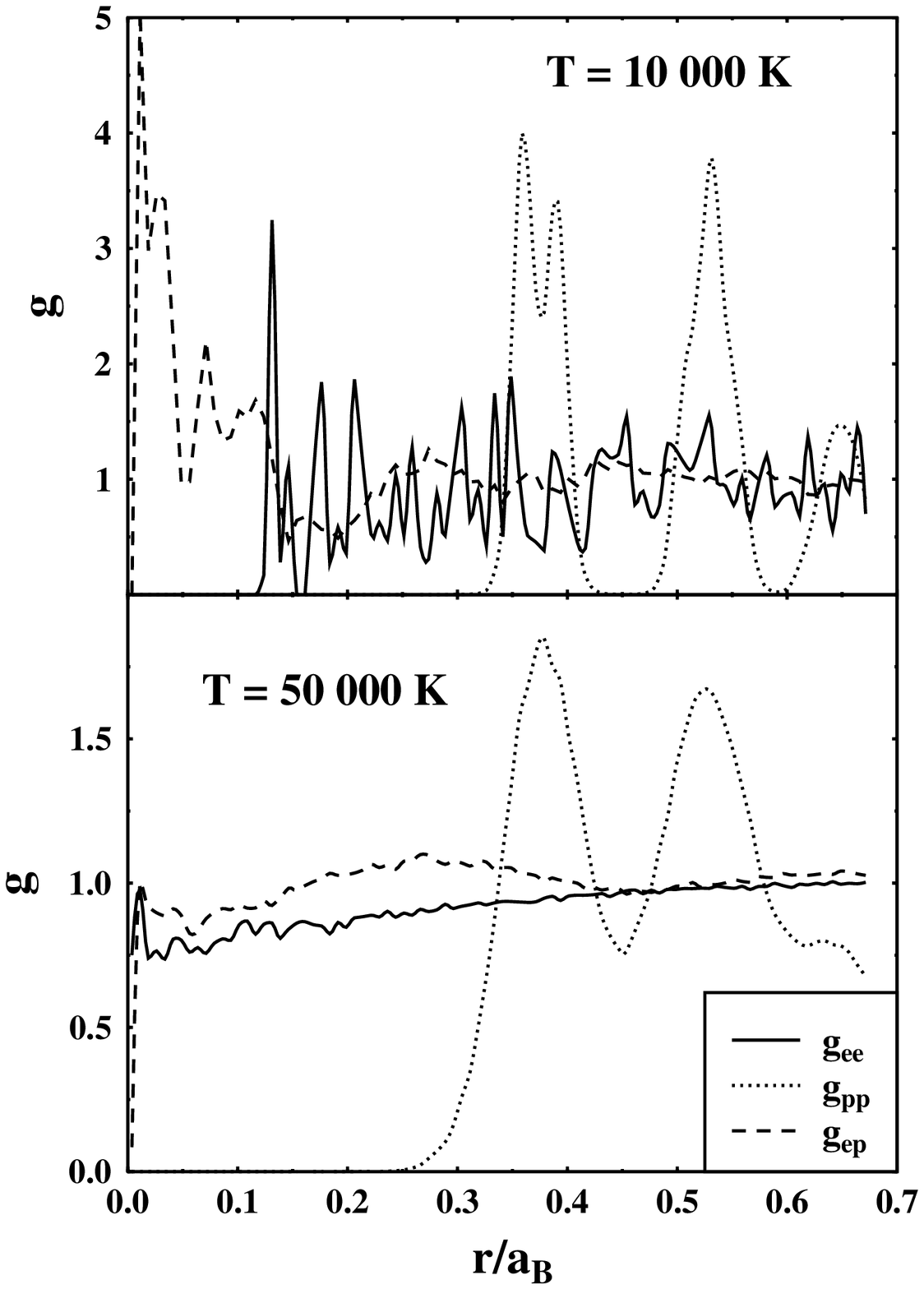,height=18cm}}
\vspace{-1cm}
\caption[]{\label{cp}
Electron-electron (ee), proton-proton (pp) and electron-proton (ep)
pair distribution functions of hydrogen from the PIMC simulations
at $n=10^{26}$cm$^{-3}$ for
a temperature of 10,000K (upper figure) and 50,000K (lower figure). Note
the different vertical scales.
}
\end{figure}

However, the most important effect results again from the finite-size character of our 
simulations. To better understand the high-density results, we analyze in Fig.~\ref{cp} the 
electron-electron (e-e), proton-proton (p-p) and electron-proton (e-p) pair distribution 
functions. These functions exhibit features typical for strongly correlated systems.
The most prominent effect is seen in the p-p function which exhibits a periodic 
structure at $T=50,000$K which is even more pronounced at $T=10,000$K. This proton ordering
is typical for a strongly correlated ion fluid which is near the crystallization 
temperature \cite{cryst}. 
Our simulations for still higher densities show the formation of an 
ionic lattice immersed into a delocalized sea of electrons, i.e. an ionic Wigner crystal 
as it is known to exist in high density objects such as White or Brown dwarf stars. 
Thus, qualitatively, the simulations show the correct behavior at high densities. But due 
to the small size of the simulations (only 50 electrons and protons are presently feasible),
the results are much closer to those for small ionic clusters which are known to exhibit quite peculiar
behavior, including strong size dependence of the energy, negative specific heat etc. 
Therefore, in order to obtain more accurate data for the internal energy of a macroscopic 
two-component plasma at ultrahigh compression, a significant increase of the simulation size is
desirable which should become feasible in the near future.

\section{Discussion}\label{dis}

This work is devoted to the investigation of the thermodynamic properties of
hot dense plasmas in the temperature region between $10,0000$ and $50,000$K.
We presented a new theoretical approach to high-density plasmas which is based on the 
theory of an effective ion-ion potential (EIIP). This method is shown to be quite 
efficient for fully ionized strongly correlated plasmas above the Mott density.

Furthermore, a detailed comparison of several theoretical approaches on one hand and 
simulations on the other, has been performed over a wide density range. The first 
include the analytical models EEIP and the PACH on one hand and recent density
functional data of Xu and Hansen \cite{Xu98} on the other hand. 
The second group of data includes several new data points based on direct path integral
Monte Carlo simulations (PIMC) of a correlated proton-electron system with degenerate 
electrons. In addition, we compared with restricted PIMC data of Militzer et al. 
\cite{militzer-etal.00}.
 
From this comparison we conclude that the three theoretical approaches are in 
very good agreement with each other for a fully ionized hydrogen plasma in the 
high density region where $r_s<1$. On the other hand, the two simulations agree 
with each other for temperatures above $50,000$K although no RPIMC data for 
high densities are yet available to us. This agreement over a broad range of parameters 
is certainly remarkable since the plasma is far outside the perturbative regime: it is 
strongly correlated and the electrons are degenerate. Moreover, all considered 
methods are essentially independent.

Finally, the comparison of our PIMC simulation results with the analytical data 
reveals an overall good agreement, although deviations are observed above 
$n=10^{22}$cm$^{-3}$. The simulation energy reaches a far deeper minimum and the 
energy increase due to electron degeneracy appears at higher densities. The 
discrepancy at lower densities (below $n=10^{23}$cm$^{-3}$) was attributed to not 
adequate treatment of bound state effects in the analytical methods, whereas 
the deviations at higher density are most likely due to finite size effects encountered
by the simulations. These lead to droplet formation at low temperature and for 
densities between $n=10^{23}$cm$^{-3}$ and $n=10^{24}$cm$^{-3}$ which are an 
indication for the plasma phase transition \cite{filinov-etal.01jetpl}. At high 
density, the simulations reveal ordering of protons into a strongly correlated
fluid and onset of the formation of a proton Wigner crystal. 
These interesting physical effects in high pressure hydrogen are of relevance 
for many astrophysical systems, but also for many laboratory experiments, including
ultracold degenerate trapped ions and laser plasmas.

\section{Acknowledgements}
We acknowledge stimulating discussions with
W.D.~Kraeft, D.~Kremp, R. Redmer and M.~Schlanges on the properties of 
hydrogen.

This work has been supported by the Deutsche
Forschungsgemeinschaft (grant BO-1366/2) and by a grant for
CPU time at the NIC J\"ulich.

\begin{references}
\bibitem{cryst} In fact, the first minimum of the proton-proton function (around $r=0.45a_B$)
for $T=10,000$K is far lower than the standard value of 0.35 typical for a liquid.
\end{references}
\end{document}